\documentclass[twocolumn]{svjour3}          
\smartqed  
\usepackage{graphicx}
\usepackage{subfigure}
\usepackage{algorithm}
\usepackage{algorithmic}
\usepackage{amsmath}
\usepackage{amssymb}
\usepackage{multirow}
\journalname{XXX}
\begin{document}
	
	\title{Personalized QoS Prediction of Cloud Services via Learning Neighborhood-based Model}
	
	\author{Hao Wu*
		\and Jun He
		\and Bo Li
		\and Yijian Pei
	}
	
	
	\institute{H. Wu,   B. Li,   Y. Pei \at
		School of Information Science  and Engineering,   Yunnan University,   Kunming 650091,   China \\
		Tel.: +86-0871-65033748\\
		Fax: +86-0871-65033748\\
		\email{\{haowu,   libo,   pei3p\}@ynu.edu.cn}
		\and
		J. He \at
		Nanjing University of Information Science and Technology,   Nanjing 210044,   China\\
		\email{hejun.zz@gmail.com}
		\\
		*Co-corresponding Author
	}
	
	\date{Received: date / Accepted: date}

	\maketitle
	
	\begin{abstract}
		The explosion of cloud services on the Internet brings new challenges in service discovery and selection. Particularly, the demand for efficient quality-of-service (QoS) evaluation is becoming urgently strong. To address this issue, this paper proposes neighborhood-based approach for QoS prediction of cloud services by taking advantages of collaborative intelligence. Different from heuristic collaborative filtering and matrix factorization, we define a formal neighborhood-based prediction framework which allows an efficient global optimization scheme, and then exploit different baseline estimate component to improve predictive performance. To validate the proposed methods, a large-scale QoS-specific dataset which consists of invocation records from 339 service users on 5,825 web services on a world-scale distributed network is used. Experimental results demonstrate that the learned neighborhood-based models can overcome existing difficulties of heuristic collaborative filtering methods and achieve superior performance than state-of-the-art prediction methods.
	\end{abstract}
	\section{Introduction}
	Cloud computing provides users with transparent and elastic infrastructure to deploy various applications (e.g. e-mail, multimedia, social network, e-commerce applications) and allow ubiquitous network to access these applications. From a service-oriented architecture perspective \cite{Barry2013}, cloud services refer to those applications provided to users on demand via the Internet from a cloud computing provider's servers. The explosion of cloud services on the Internet brings new challenges in service discovery and selection. To address these challenges, a range of studies have been carried out to develop advanced techniques that will assist users to choose appropriate services  \cite{Sun2014}. \par
	Existing techniques,  such as web service retrieval \cite{Hao2010} and semantic service discovery \cite{Ngan2013},  can be used to locate services satisfying users' required functionality. Nevertheless,  as confronted with a number of functionally similar cloud services,  a user may feel hard to judge what is the extent of candidate services in line with the individual needs. A further comparison of the non-functionality of candidate services (generally the properties of quality of services) is necessary in order to make the best choice. To obtain accurate and personalized client-side quality values for individual users, client-side service evaluations \cite{Deora2003,Wu2007,Tsai2008} are usually needed. However, due to constraints on time, costs and other factors \cite{Zheng2013}, service providers cannot dispose a large number of software sensors in cloud environments to monitor QoS information for every service. Also, it is not realistic for users to carry out large-scale testings to experience the individual differences of QoS. Consequently, how to obtain personalized QoS of cloud services and assist users selecting appropriate services become one urgent issue. \par
	Recently,  researchers from academia and industry begin to pursuit solution for this problem by drawing lessons from the recommender systems \cite{Shao2007,Zheng2009,Zheng2011,Chen2014,Wu2013,Sun2013,Zheng2010,Zheng2013,ZhangRC2014,ZhangWC2014,Lo2015,Yin2014,Yu2013} . The main idea of their works is to analyze the QoS usage data of users in service-oriented systems,  further exploit collaborative intelligence \cite{Gill2012} to prediction unknown QoS value \cite{Shao2007,Zheng2009,Zheng2010}. For example,  given a target service $i$ and an active user $u$,  the historical quality records of the target service $i$ have being observed by the top-k similar neighbors of the current user $u$ can be synthesized and taken as the predicted value of $u$ to $i$ \cite{Shao2007}. With distinguishable quality values of candidate services,  users can take a decision on choosing appropriate services. In such a manner, it can avoid direct QoS measurement,  and thereby save time and economic costs for both service providers and users. \par
	With respect to the collaborative QoS prediction,  the commonly used methods are neighborhood-based collaborative filtering (CF) \cite{Adomavicius2005} and matrix-factorization \cite{Koren2009}. The advantages of neighborhood-based CF are simplicity, justifiability and efficiency \cite{Desrosiers2011}. However,  these models are not justified by a formal model. Moreover, heterogenous similarity metrics and sparsity-sensitive problem make these models not robust and scalable enough.  In contrast, matrix-factorization approaches comprise an alternative approach to CF with the more holistic goal to uncover latent features from user-service usage data. Thus can alleviate the problem of sensitivity to sparse data. Since matrix-factorization can be presented as a formal optimization problem and solved by machine learning methods, thus it provides attractive accuracy and scalability for QoS prediction. However, matrix-factorization is always uncertain, resulting in difficulty as explain the predictions for users. It is very important that recommender systems provide explanations for their recommendations so that users can consider and trust them \cite{Cleger2014}. By the same token, users of cloud services certainly hope to get reasonable explanations for the QoS predictions provided by a service recommendation system. \par
	Consequently, a spontaneous concern may be raised whether we can develop more accurate neighborhood models which overcome existing difficulties, and achieve more accurate prediction than matrix factorization. It would be a better solution for the task of QoS prediction.   Inspired by this, we propose learning neighborhood-based models for personalized QoS prediction of cloud services. We define formal neighborhood models which permit an efficient global optimization scheme and exploit different baseline estimate components to improve prediction performance. Experimental results show that learning neighborhood model can overcome existing difficulties, and perform superior to the-state-of-art prediction methods.  \par
	The remainder of the paper is organized as follows. We review some existing works that are most relevant to ours in Section 2. We give details of the proposed neighborhood-based prediction approaches in Section 3. We measure the effectiveness of the proposed methods via a set of experiments on real QoS data in Section 4 and conclude in Section 5.
	\section{Related works}
	\subsection{QoS prediction based on neighborhood model}
	The most common approach to CF-based QoS prediction is neighborhood-based models which carry on similar users or services to predict the QoS of target service. It can be divided into user-based  and service-based collaborative prediction model. One basic user-specific model can be formulated as following,  
	\begin{equation}\label{eq:1}
	\hat{r}_{ui}=\mu_u+\frac{\sum_{v\in \mathcal{N}^k_u} S_{uv} (r_{vi}-\mu_v)}{\sum_{v\in \mathcal{N}^k_u} S_{uv}}
	\end{equation}
	where,  $ \mathcal{N}^k_u$ is the set of k-nearest neighbors of user $u$,  $\mu_u$ and $\mu_v$ are the average QoS value observed by user $u$ and user $v$. $S_{uv}$ is the similarity between $u$ and $v$ and can be described with numerous metrics,  such as pearson correlation coefficient (PCC),  cosine similarity,  and mean squared difference \cite{Desrosiers2011}.  To implement service-based prediction model, we can just switch the roles of users and services. \par
	Shao et al. \cite{Shao2007} at first proposed the use of collaborative prediction method based on the user. Firstly,  PCC measurement is utilized to calculate pairwise-similarity among all users on the user-service matrix of QoS data. Secondly, historical quality values of target service provided by the top-k similar users of active user are fused to achieve prediction result.  Follow-up research works basically follow this idea but concentrate on improving the similarity metrics to accurately quantify the correlations of users or services. Zheng et al. \cite{Zheng2009,Zheng2011} proposed a mixed model that integrated user-based and item-based approaches linearly by confidence weights and proved the mixed model is better than a single one. Sun et al. considered the distribution characteristics of QoS data to calculate the similarity \cite{Sun2013}. Chen et al. \cite{Chen2014} and Wu et al. \cite{Wu2013}  proposed location-aware similarity metrics to find neighbors of users and services. These neighborhood-based methods do not exploit machine learning, hence, can be seen as heuristic-based prediction approaches. \par 
	Neighborhood-based methods became very popular because they are relatively simple to implement and provide intuitive explanations for the prediction results. However, some concerns about neighborhood-based methods always exist. Most notably, these methods are not justified by a formal model. The selection of heterogenous similarity metric clearly affects the accuracy of QoS prediction, thus make these models not robust. When the QoS data are sparse, the predictive power can be greatly reduced, resulting in the sparsity-sensitive problem. This motivates us to develop more accurate neighborhood models to resolve existing difficulties.
	\subsection{QoS prediction based on matrix factorization}
	Different from collaborative filtering,  matrix factorization approaches have attractive accuracy and scalability thus recently become popular in recommender systems \cite{Koren2010}. A typical model associates each user $u$ with an user-factors vector $p_u\in R^f$ ,  and each item $i$ with an item-factors vector $q_i\in R^f$ . The prediction is done by taking an inner product,  i.e.,  $\hat{r}_{ui}=p_u^Tq_i$. To exploit matrix factorization for QoS prediction,  Zheng et al. \cite{Zheng2010,Zheng2013} using probabilistic matrix factorization (PMF \cite{Salakhutdinov2007}) approach to decompose the QoS matrix. For identifying latent factors,  $p$'s and $q$'s,  a least-square's problem like Eq.\ref{eq:2} are built and solved using gradient descent \cite{Koren2010} .
	\begin{equation}
	\label{eq:2} 
	\min_{p*, q*} \sum_{ (u, i)\in \mathbf{E}} (r_{ui}-p_u^Tq_i)^2+\lambda_u||p_u||^2+\lambda_v  ||q_i||^2 
	\end{equation}\par 
	Matrix factorization can partially alleviate sparity-sensitive problem of collaborative filtering. Thus improve the accuracy of QoS prediction. For the last two years, numerous efforts have been made on improving MF-based models. These works concentrate on utilizing of additional information, such as spatial and temporal information associated with users or services. Zhang et al. \cite{ZhangRC2014} use collective matrix factorization that simultaneously factor the user-service quality matrix, service category and location context matrices. Zhang et al. \cite{ZhangWC2014} factorize user-service-time matrix of QoS using non-negative tensor factorization with time information. Yin et al. \cite{Yin2014} develop a location-based regularization framework for PMF prediction model. Lo et al \cite{Lo2015} exploit PMF prediction model with a localtion-based pre-filtering stage on QoS matrix. He et al. \cite{He2014} develop location-based hierarchical matrix factorization. Yu et al. \cite{Yu2013} experience trace-norm regularized matrix factorization. \par
	Matrix factorization is still uncertain, resulting in difficulty as explain the predictions for users. Also, data sparsity has a negative impact on these methods, as data becomes extremely sparse,  the prediction performance will be not optimistic.
	\section{Learning Neighborhood-based Model for QoS Prediction} 
	\subsection{Problem Description}
	To cope with existing drawbacks of CF-based prediction methods,  we suggest using machine learning techniques to build neighborhood model for QoS prediction. New models allow an efficient global optimization scheme and exploit different baseline estimate component to improve prediction accuracy. We reserve special indexing letters for distinguishing users from service: for users $u$,  $v$,  and for services $i$,  $j$. A QoS $r_{ui}$ indicates the observed quality of user $u$ on service $i$. We distinguish predicted quality from known ones,  by using the notation $\hat{r}_{ui}$ for the predicted value of $r_{ui}$. The $ (u,  i)$ pairs for which $r_{ui}$ is known are stored in the set $\mathbf{E} = \{ (u,  i) | r_{ui} \quad is  \quad known\}$. Usually the vast majority of QoS values are unknown. To combat overfitting in learning prediction model on the sparse data,  models are regularized so estimates are shrunk towards baseline defaults. \par
	\subsection{Neighborhood models}
	User-oriented methods estimate unknown quality based on recorded QoS of like minded users. Analogously,  in service-oriented methods,  a QoS is estimated using known QoS made by the same user on similar services. In cloud computing,  the context with users is more complicated and dynamic than that of services.
	Prediction leveraged by similar users other than services is more reasonable.  Thus,  our focus is on user-oriented approaches,  but parallel techniques can be developed in a service-oriented fashion,  by switching the roles of users and services.\par
	We exploit the neighborhood model proposed in collaborative filtering research \cite{Koren2010}, which allows an efficient global optimization scheme and offers improved accuracy. To facilitate global optimization, we would wish to abandon such user-specific weights (see $S_uv$ in Eq.\ref{eq:1}) in favor of global weights independent of a specific user. The weight from user $v$ to user $u$ is denoted by $w_{uv}$ are able to be learned from the data through optimization. By this, we can overcome the weaknesses with existing neighborhood-based models.  An initial sketch of the model describes each quality score $r_{ui}$ by Eq.\ref{eq:3}:
	\begin{equation}\label{eq:3}
	\hat{r}_{ui}=b_{ui}+\sum_{v\in \mathcal{N}_u} (r_{vi}-b_{vi})w_{uv}, 
	\end{equation}
	where $\mathcal{N}_u$ is the neighbor set of user $u$,  $b_{ui}$ is the basic estimate that we will gradually construct considering different factors.\par
	With respect to the interpretation of weights,  usually they represent interpolation coefficients relating unknown quality score to the existing ones in a traditional neighborhood model (recall $S_{uv}$ in Eq.\ref{eq:1}). Here, we adopt them in a different viewpoint that weights represent offsets to basic estimates and residual, $r_{vi}-b_{vi}$, is viewed as the coefficients multiplying those offsets. For two similar users $u$ and $v$, $w_{uv}$ is always expected to get high, and verse visa. So, our estimate will not deviate much from the basic estimate by a user $v$ that accessed $i$ just as expected ($r_{vi}-b_{vi}$ is around zero),  or by a user $v$ that is not known to be predictive on $u$ ($w_{uv}$ is close to zero).\par
	Generally, we can take all users in $\mathcal{N}_u$ other than $u$,  however, this would increase the number of weights to be estimated.  In order to reduce complexity of the model, we suggest pruning parameters corresponding to unlikely user-user relations. Let $\mathcal{N}^k_u $ be the set of $k$ users most similar $u$, as determined by the similarity measure $S_{uv}$ (see Eq.\ref{eq:1}). Further,  we let $ \mathcal{N}^k_{(i;u)} \triangleq   \mathcal{N}_i \cap \mathcal{N}^k_{u} $, where $\mathcal{N}_i$ is the set of users have used the service $i$.  Now, when predicting $r_{ui}$ according to formula \ref{eq:3}, it is expected that the most influential weights will be associated with users similar to $u$. Hence, we replace Eq.\ref{eq:3} with:
	\begin{equation}\label{eq:4}
	\hat{r}_{ui}=b_{ui}+ |\mathcal{N}^k_{(i;u)}|^{-\frac{1}{2}}\sum_{v\in \mathcal{N}^k_{(i;u)}} (r_{vi}-b_{vi})w_{uv}
	\end{equation}
	When $k = \infty$,  rule (\ref{eq:4}) coincides with (\ref{eq:3}). When $k = 0$, $\hat{r}_{ui}=b_{ui}$. However, for other values of $k$, it offers the potential to significantly decrease the number of variables involved. This final prediction rule permits fast online prediction, since more computational works, such as similarity calculation and parameter estimation, have been made at the pre-processing stage.  Recall that unlike matrix factorization, the neighborhood models allow a direct explanation of their predictions, and do not require re-training the model for handling new services. 
	\begin{figure*}[ht]
		\centering
		\includegraphics[width=6in]{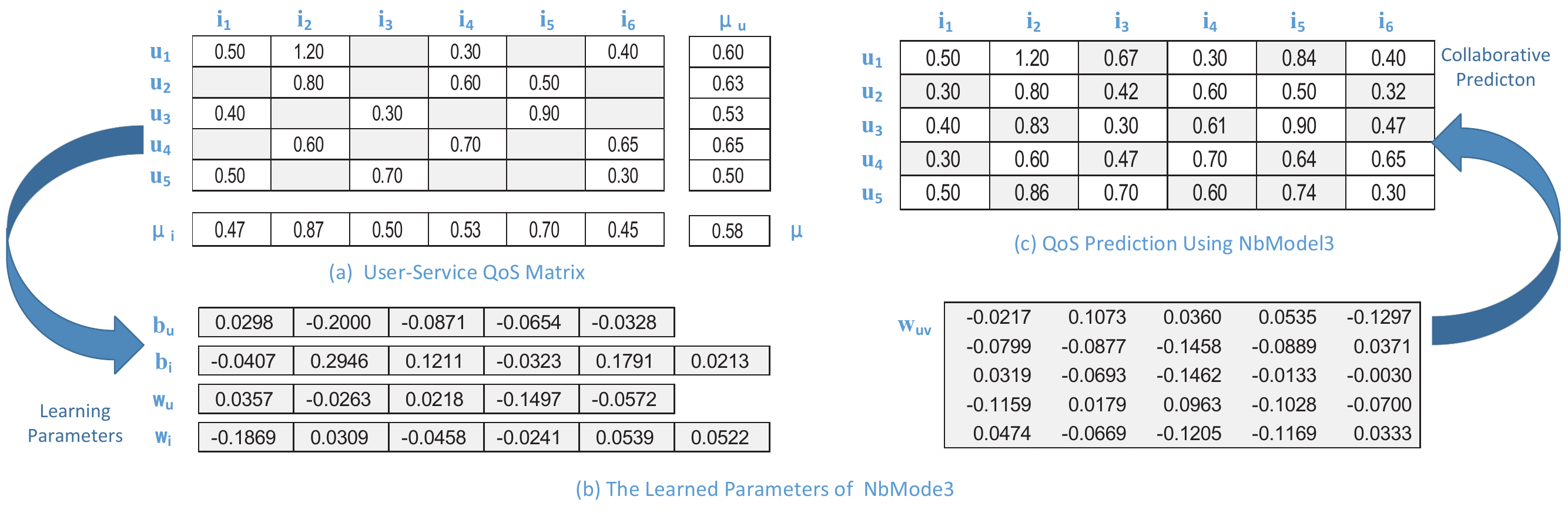}
		\caption{\label{fig:1} An example on QoS prediction with learning neighborhood-based prediction model.}
	\end{figure*}
	\subsection{Components for estimating $b_{ui}$}
	In this section,  we will gradually construct the component to estimate $b_{ui}$ through an ongoing refinement of our formulations. 
	\paragraph{Baseline estimate.} Typical QoS data exhibit large user and service effects-i.e.,  systematic
	tendencies for some users to achieve better QoS than others, and for some services to receive better QoS than others. It is usual to adjust the QoS data by accounting for these effects, which we encapsulate within the baseline estimates. Denote by $\mu$ the average QoS value observed in the entire dataset. A baseline estimate for an unknown QoS $r_{ui}$ is denoted by $b_{ui}$ and accounts for the user and service effects:
	\begin{equation}\label{eq:5}
	b_{ui} = \mu +b_u+b_i
	\end{equation}
	The parameters $b_u$ and $b_i$ indicate the observed deviations of user $u$ and service $i$,  respectively, from the average. For instance, suppose that we want to find out the baseline estimate for the response time of Google-Search service by Tom. Now, presume that the average response-time, $\mu=2ms$. Furthermore, google-search is recognized better than an average search engine service, so we can suppose it is faster 0.5ms than the average. On the other hand, Tom's network condition is not a good, which tends to be 1ms delay than the average. Thus,  the baseline estimate for Google-Search's response-time by Tom would be 2.5ms by calculating: 2-0.5+1.  Substituting Eq.\ref{eq:5} into Eq.\ref{eq:4}, we will obtain the first neighborhood model (NbModel1). 
	To estimate $b_u$,  $b_i$ and $w_{uv}$ one can solve the least squares problem:
	\begin{equation}\label{eq:6}
	\begin{aligned}
	\min_{w*, b*} &\sum_{ (u, i)\in \mathbf{E}}[r_{ui}-b_{ui}-|\mathcal{N}^k_{(i;u)}|^{-\frac{1}{2}}\sum_{v\in \mathcal{N}^k_{(i;u)}} (r_{vi}-b_{vi})w_{uv}]^2\\
	&+\lambda_1\sum_{v\in \mathcal{N}^k_{(i;u)}}w_{uv}^2+\lambda_2 (b_u^2 +b_i^2)
	\end{aligned}
	\end{equation}
	Here, the first term $\sum_{ (u, i)\in \mathbf{E}} (.)^2$ strives to find
	$b_u$'s, $b_i$'s and $w_{uv}$'s that fit the given response data. The second part and the third part both are the regularizing terms, used to avoid overfitting by penalizing the magnitudes of the parameters. $\lambda_1$ and $\lambda_2$ are the specific regularization parameters \cite{Koren2009,Koren2010}.
	\paragraph{Weighted features.}
	The baseline estimate just considers the mean effect of user and item in an intuitive manner. However, more bias information (such as user features, item features and time bias information, etc) can be utilized to enhance the prediction model. We may consider an unknown quality value $r_{ui}$ as a linear combination of the features of user and item. Here, the user-specific QoS mean and the service-specific QoS mean are taken as the two key features,  because that this two are strong features to reflect bias effect of users and services,  and the expect prediction of QoS can be found in a value domain determined by them.  Denote by $\mu_u$ the average QoS value observed by user $u$ and $\mu_i$ the average QoS value observed by service $i$. A feature-weighted estimate for $r_{ui}$ is denoted by $b_{ui}$ as followings:
	\begin{equation} \label{eq:7}
	b_{ui} = w_u \mu_u +w_i  \mu_i
	\end{equation}
	The parameters $w_u$ and $w_i$ indicate the feature importance of user $u$ and item $i$, respectively. Substituting Eq.\ref{eq:7} into Eq.\ref{eq:4}, we will obtain the second neighborhood model (NbModel2). In order to estimate $w_u$, $w_i$, we need to solve the following least squares problem:
	\begin{equation}\label{eq:8}
	\begin{aligned}
	\min_{w*} &\sum_{ (u, i)\in \mathbf{E}}[r_{ui}-b_{ui}-|\mathcal{N}^k_{(i;u)}|^{-\frac{1}{2}}\sum_{v\in \mathcal{N}^k_{(i;u)}} (r_{vi}-b_{vi})w_{uv}]^2\\
	&+\lambda_1\sum_{v\in \mathcal{N}^k_{(i;u)}}w_{uv}^2+\lambda_3 (w_u^2 +w_i^2)
	\end{aligned}
	\end{equation}
	Here, regularizing term, $\lambda_3 (w_u^2 +w_i^2)$, avoid overfitting by penalizing the magnitudes of the parameters $w_u$'s and $w_i$'s.
	\paragraph{Hybrid approach.}
	Beyond estimating $b_{ui}$ based on either the baseline estimate or the weighted features,  we may combine them together to have both of worlds. This leads to a new prediction rule for $b_{ui}$:
	\begin{equation}\label{eq:9}
	b_{ui} =\mu +b_u+b_i+ w_u \mu_u +w_i  \mu_i
	\end{equation}
	Substituting Eq.\ref{eq:9} into Eq.\ref{eq:4}, we will obtain the third neighborhood model (NbModel3). To estimate $w_u$, $w_i$, $b_u$, $b_i$, we need to solve the regularized least squares problem as followings:
	\begin{equation}\label{eq:10}
	\begin{aligned}
	\min_{w*, b*} &\sum_{ (u, i)\in \mathbf{E}} [r_{ui}-b_{ui}-|\mathcal{N}^k_{(i;u)}|^{-\frac{1}{2}}\sum_{v\in \mathcal{N}^k_{(i;u)}} (r_{vi}-b_{vi})w_{uv}] ^2\\
	&+\lambda_1\sum_{v\in \mathcal{N}^k_{(i;u)}}w_{uv}^2+\lambda_2 (b_u^2 +b_i^2)+\lambda_3 (w_u^2 +w_i^2)
	\end{aligned}
	\end{equation}
	\subsection{Models Learning}
	In a sense, our neighborhood models provide two-tier models for personalized QoS prediction. The first tier, $b_{ui}$, describes general properties of the service and the user, without accounting for any involved interactions. The second tie-"Neighborhood tier" contributes fine grained adjustments that are hard to profile. Model parameters are determined by minimizing the associated regularized squared error function through gradient descent. Recall that $e_{ui} \triangleq r_{ui}-\hat{r}_{ui}$. We loop over all known ratings in $\mathcal{E}$. For a given training case $r_{ui}$, we modify the parameters by moving in the opposite direction of the gradient, yielding:
	\begin{itemize}
		\item $b_u \leftarrow b_u+\gamma_1  (e_{ui} -\lambda_2  b_u)$
		\item $b_i \leftarrow b_i+\gamma_1  (e_{ui} -\lambda_2  b_i)$
		\item $w_u \leftarrow w_u+\gamma_1  (e_{ui}  \mu_u-\lambda_3  w_u)$
		\item $w_i \leftarrow w_i+\gamma_1  (e_{ui}  \mu_i-\lambda_3  w_i)$
		\item $\forall v \in \mathcal{N}^k_{(i;u)}$ \par
		$w_{uv} \leftarrow w_{uv}+ \gamma_2  \left (  |\mathcal{N}^k_{(i;u)}|^{-\frac{1}{2}}  e_{ui}  (r_{vi}-b_{vi})- \lambda_1  w_{uv}\right) $
	\end{itemize}
	Note that, update rules set forth can fit all of the least squares problems in Eqs.\ref{eq:6}, ref{eq:8} and \ref{eq:10}. When assessing the method on a given dataset, we took advantage of following values for the meta parameters: $\lambda_1=\lambda_2=\lambda_3=0.001$, $\gamma_1=\gamma_2=0.001$.  It is beneficial to decrease step sizes (the $\gamma$'s) by a factor of 0.9 after each iteration. Another important parameter is $k$, which controls the neighborhood size. Our experience shows that increasing $k$ always benefits the RMSE of the results on the test set. Hence, the choice of $k$ should reflect a tradeoff between prediction accuracy and computational cost. A toy-example on Qos prediction using NbModel3 is given in Figure \ref{fig:1}, where we let $k=5$.
	\section{Experiments}
	\subsection{Datasets}
	To evaluate the QoS prediction performance, we use a large-scale dataset collected by Zheng et al.\cite{Zheng2013,Zheng20014a}.
	The dataset consists of a total of 1, 974, 675 real-world web service invocation results are
	collected from 339 users on 5, 825 real-world web services via PlanetLab platform \footnote{http://planet-lab.org/},  
	which is a global research network that supports the development of new network services.
	This dataset can be treated as a set of usage data for real-world cloud services from distributed locations.
	For more details, users can refer to works \cite{Zheng2013,Zheng20014a}. In our experiments, we only think the response time (the range scale is 0\~20s ). However, the proposed approach can be applied to other QoS properties easily.
	\begin{table*}[ht]
		\centering
		\caption{Performance comparisons of QoS prediction models using different matrix density,  where '*' and '**' indicate 1st-class and 2nd-class,  respectively.}
		\label{table:1}
		\begin{tabular}{c|ll|ll|ll|ll}
			\hline
			\multirow{2}{1.5cm}{\textbf{Methods}} &\multicolumn{2}{c|}{\textbf{MD}=0.5\%} &\multicolumn{2}{c|}{\textbf{MD}=1\%}  &\multicolumn{2}{c|}{\textbf{MD}=5\%} &\multicolumn{2}{c}{\textbf{MD}=10\%} \\
			\cline{2-9}
			&\textbf{MAE} &\textbf{RMSE}  &\textbf{MAE} &\textbf{RMSE}  &\textbf{MAE} &\textbf{RMSE}  &\textbf{MAE} &\textbf{RMSE} \\
			\hline
			\textbf{GMEAN}   		&0.9721 &1.9735 &0.9915 &1.9726 &0.9860 &1.9740 &0.9920 &1.9725 \\
			\textbf{UMEAN}   		&0.8873 &1.8914 &0.8899 &1.8705 &0.8746 &1.8600 &0.8744 &1.8575\\
			\textbf{IMEAN}   		&0.8317 &1.9326 &0.7870 &1.8136 &0.7002 &1.5746 &0.6890 &1.5410\\
			\textbf{UPCC}    		&0.9709 &1.9727 &0.9468 &1.9457 &0.6173 &1.3925** &0.5446 &1.3119\\
			\textbf{IPCC}    		&0.9721 &1.9735 &0.9888 &1.9716 &0.6675 &1.4272 &0.6430 &1.3798\\
			\textbf{UIPCC}    	    &0.9708 &1.9725 &0.9453 &1.9430 &0.6162 &1.3900** &0.5439 &1.3102\\
			\textbf{PMF}    		&0.8317 &2.0624 &0.8195 &2.0010 &0.6354 &1.5071 &0.5541 &1.3393\\
			\textbf{NMF}    		&0.7656* &1.8480 &0.7092* &1.7650 &0.6516 &1.5145 &0.6461 &1.4527\\
			\textbf{BiasedMF}     &0.8052 &1.7643 &0.7832 &1.7128 &0.6376 &1.4361 &0.5518 &1.3088\\
			\textbf{NbModel1}     &0.7986 &1.6989* &0.7471 &1.6163* &0.6106 &1.5333 &0.5336 &1.3563\\
			\textbf{NbModel2}     &0.7678* &1.7942 &0.7196** &1.6436** &0.5817** &1.3943** &0.5156** &1.2832**\\
			\textbf{NbModel3}     &0.7838** &1.7847** &0.7428 &1.6582 &0.5793* &1.3795* &0.5138* &1.2739*\\
			\hline
		\end{tabular}
	\end{table*}
	\subsection{Evaluation metrics}
	Mean absolute error (MAE) and root mean squared error (RMSE) metrics,  two basic statistical accuracy metrics \cite{Hyndman2006}, have been extensively used in  performance evaluation of rating predictions \cite{Adomavicius2005}, are used to measure the QoS prediction performance of selected methods.
	MAE is defined as 
	\begin{equation}\label{eq:11}
	MAE=\frac{\sum_{u, i}|r_{ui}-\hat{r}_{ui}|}{N}
	\end{equation}
	where $r_{ui}$ is the observed QoS value,  $\hat{r}_{ui}$ is the predicted one,  and $N$ is
	the number of test cases. The MAE measures the average magnitude of the errors in a set of forecasts, without considering their direction \cite{Hyndman2006}.	RMSE is defined as
	\begin{equation}\label{eq:12}
	RMSE=\sqrt{\frac{\sum_{u, i} (r_{ui}-\hat{r}_{ui})^2}{N}}
	\end{equation}
	The RMSE is a quadratic scoring rule which measures the average magnitude of the error. \par
	Both MAE and RMSE, a smaller value indicates a better prediction accuracy. The MAE is a linear score which means all the individual differences are weighted equally in the average. Since the errors are squared before they are averaged, the RMSE gives a relatively high weight to large errors. This means the RMSE is most useful when large errors are particularly undesirable.
	\subsection{Comparison}
	To show the prediction accuracy of our neighborhood-based  approaches,  we compare our methods with the three kinds of popular approaches:
	\begin{enumerate}
		\item \textit{Statistical approaches }
		\begin{enumerate} 
			\item \textit{GMEAN} takes the average QoS value of whole dataset as the predictive QoS value of user $u$ to service $i$,  i.e. $\hat{r}_{ui}=\mu$; 
			\item \textit{UMEAN} takes the average QoS value known by $u$ as the predictive QoS value of $u$ to $i$ ,  i.e., $\hat{r}_{ui}=\mu_u$;
			\item \textit{IMEAN} takes the average QoS value observed from $i$ as the predictive QoS value of $u$ to $i$,  i.e. $\hat{r}_{ui}=\mu_i$; 
		\end{enumerate}
		\item \textit{Heurstic-based CF}
		\begin{enumerate}
			\item \textit{UPCC} is user-based collaborative prediction model. Top-k neighbors of users are found using PCC-based similarity \cite{Shao2007}; 
			\item \textit{IPCC} is item-based collaborative prediction model. Top-k neighbors of items (services) are found using PCC-based similarity \cite{Zheng2011};
			\item \textit{UIPCC} combines the user-based and item-based collaborative prediction approaches and employs both the similar users and similar services for the QoS value prediction \cite{Zheng2011}. 
		\end{enumerate}
		\item \textit{MF-based approaches}
		\begin{enumerate}
			\item \textit{PMF} uses probabilistic matrix factorization \cite{Salakhutdinov2007} to factorize user-service QoS matrix for the prediction \cite{Zheng2013}; 
			\item \textit{NMF} uses non-negative matrix factorization \cite{Lee2000} to factorize the QoS matrix into two matrices $p$ and $q$,  with the property that all three matrices have no negative elements; 
			\item \textit{BiasedMF} exploits a combination of baseline estimate (same to Eq.\ref{eq:5}) and matrix factorization prediction rule for collaborative filtering \cite{Koren2009}. We adopt it for the QoS prediction.
		\end{enumerate}
	\end{enumerate}\par
	For the memory-based CF methods, we chose the neighborhood size of users at $k=10$ and services at $k=50$. For the matrix-factorization based methods, the regularization parameters for user and service are set at $\lambda_u=\lambda_v=0.001$, and the dimensionality of latent factors is fixed at 10.  For all selected methods, we use their implementations in LibRec \footnote{http://www.librec.net/index.html}, which provides a Java library for recommender systems and a chance for reproduction of experimental results. Note that,  for a fair comparison of selected methods,  all of them exploit only the information supplied by the user-service QoS matrix,  and no additional information (e.g.,  geo-locations of users) is allowed (we consider to extend the proposed methods to exploit additional information in the future). \par
	In the real-world, a user usually only invokes a few cloud services. Thus leading user-item matrices of QoS data sparse. This can be always observed in the recommender systems \cite{Adomavicius2005,Desrosiers2011}.  For instance, the density for the well-known Movielens1M dataset \footnote{http://grouplens.org/datasets/movielens/}
	is 4.5\%. The data density for Netflix Prize \footnote{https://en.wikipedia.org/wiki/Netflix\_prize} is about 1.4\%.  To reflect this natural phenomenon in experiments,  we randomly remove entries from the user-item matrix with different density,  specially,  we take $0.5\%-1\%$ percent for the case of sparse data and $5\%-10\%$ percent for the case of dense data.  For instance, MD (Matrix Density)=0.5\% means that we randomly select 0.5 percent of the QoS entries to predict the remaining 99.5 percent of QoS entries. The original QoS values of the removed entries are used as the expected values to study the prediction accuracy. The above-mentioned methods and neighorhood-based variants,  NbModel1,  NbModel2 and NbModel3,  are employed to forecast the QoS values of the removed entries.  The experimental results are shown in Table \ref{table:1}. \par
	Depending on Table \ref{table:1}, NbModel3 and NbModel2 respectively rank the first class and the second class on both MAE and RMSE in the case of dense data. Both models obtain smaller MAE and RMSE values consistently for response-time with MD=5\% and MD=10\%. MAE and RMSE values of neighborhood models become smaller, since denser matrix provides more information for the missing value prediction. In the case of sparse data, NMF achieve best performance in term of MAE followed by three neighborhood models. However, our methods perform much better than all other counterparts on RMSE.  Among all the prediction models, our methods achieve better performance on both MAE and RMSE, telling that learning neighborhood model can achieve higher prediction accuracy. Also, neighborhood-based models preserve the explainability of memory-based CF, and enable to give users a reason for their predictions.
	\subsection{Impact of Top-K}
	To examine the impact of top-k neighbors selection on neighborhood-based prediction models, we distinguish from two cases: sparse data and dense data. With sparse data, we found that increasing $k$ value cannot result in significant performance improvements and sometimes we may experience decreased performance. The experimental results are shown in Table \ref{table:2}, where $\hat{r}_{ui}=b_{ui}$ if  $k=0$. We believe that there are two reasons for this. On one hand, sparser matrix cannot offer the "neighborhood tier" more information to contribute fine grained adjustments. On the other hand, the leaned component $b_{ui}$ (sees Eq.\ref{eq:4}) has given ideal predictions. Nevertheless, "neighborhood ties" can be used as a regularization component to avoid the overfitting of baseline predictor even if the usage data are sparse.\par
\begin{figure*}[ht]
		\centering
		\subfigure[]{
			\label{fig:1a} 
			\includegraphics[width=2.7in]{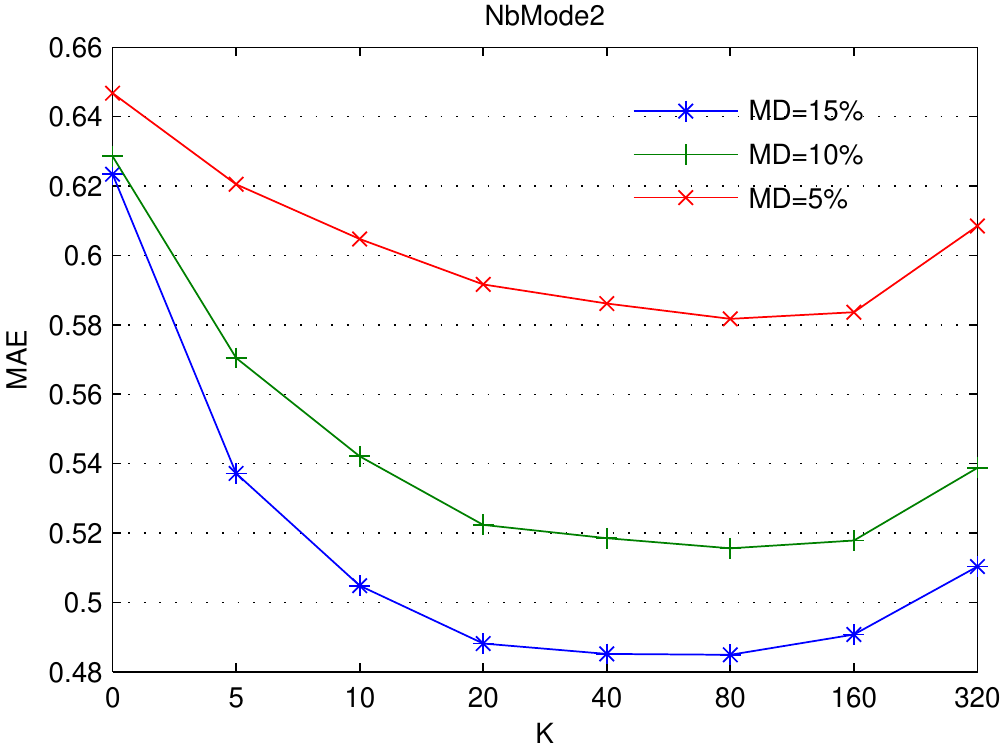}}
		\subfigure[]{
			\label{fig:1b} 
			\includegraphics[width=2.7in]{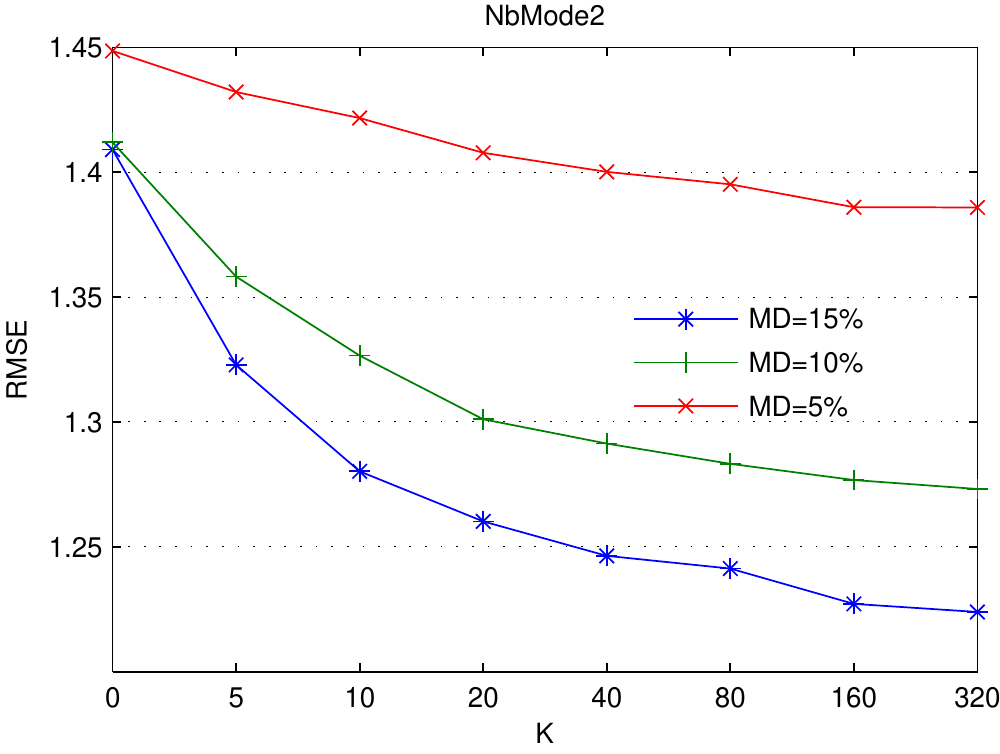}}
		\subfigure[]{
			\label{fig:1c} 
			\includegraphics[width=2.7in]{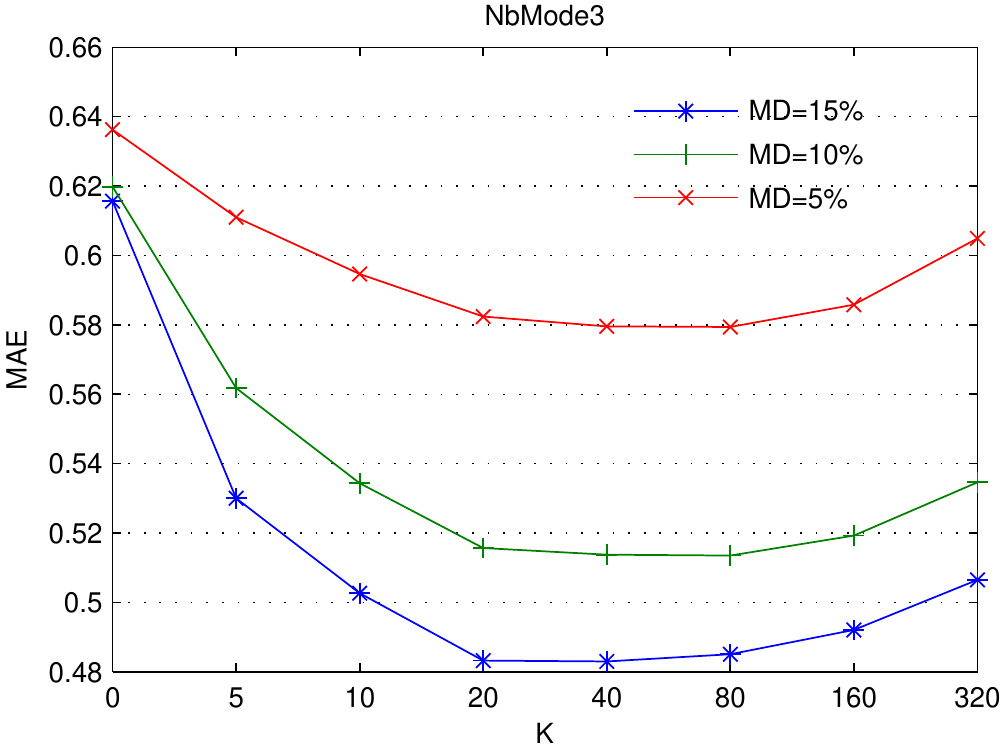}}
		\subfigure[]{
			\label{fig:1d} 
			\includegraphics[width=2.7in]{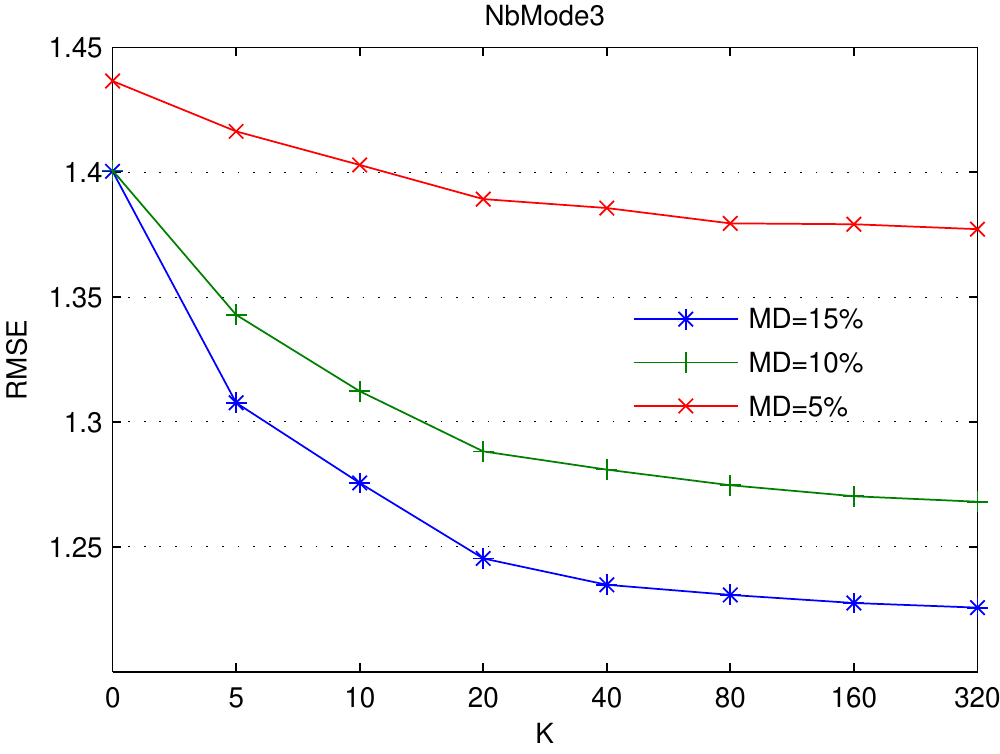}}
		\caption{\label{fig:2}Performance of NbModel2 and NbModel3 with different $k$ and matrix density.}
	\end{figure*}

	\begin{table}[ht]
		\centering
		\caption{Performance of neighbordhood-based models in case of sparse data.}
		\label{table:2}
		\begin{tabular}{c|c|ll|ll}
			\hline
			\multirow{2}{1.5cm}{\textbf{Methods}} &\multirow{2}{0.3cm}{\textbf{\textit{k}}} &\multicolumn{2}{c|}{\textbf{MD}=0.5\%} &\multicolumn{2}{c}{\textbf{MD}=1\%}  \\
			\cline{3-6} 
			&  &\textbf{MAE} &\textbf{RMSE}  &\textbf{MAE} &\textbf{RMSE} \\
			\hline
			\multirow{2}{1.5cm}{\textbf{NbModel1}}  &0     &0.7987 &1.6990 &0.7411 &1.5770\\
			&80    &0.7986 &1.6989 &0.7471 &1.6163\\
			\hline
			\multirow{2}{1.5cm}{\textbf{NbModel2}}  &0     &0.7678 &1.7942 &0.7428 &1.6567 \\
			&80     &0.7678 &1.7942 &0.7196 &1.6436 \\
			\hline
			\multirow{2}{1.5cm}{\textbf{NbModel3}}  &0     &0.7840 &1.7848 &0.7409 &1.6474 \\
			&80   &0.7838 &1.7847 &0.7428 &1.6582\\
			\hline
		\end{tabular}
	\end{table}
	In the case of condensed data, we conduct experiments to see the impact of top-K similar users based on NbModel2 and NbModel3, as both of them are more robust than NbModel1. The experimental results are shown in Figure \ref{fig:2}. From Figure \ref{fig:2}, we find that the RMSE of both models consistently decrease as increasing the value of $k$ with different matrix density (range from 5\% to 15\%). While the value distribution of MAE presents U-shaped curve, and the best configuration for our dataset is about $k=80$. In addition, more gains can be noted when the matrix becomes denser as for neighborhood-based models. For neighborhood models, since the computational cost always increases with the increment of $K$, the choice of $k$ should reflect a tradeoff between prediction accuracy and computational cost. 
	\section{Conclusion and future works}
	Based on principles of collaborative filtering and machine learning, we propose a neighborhood-based framework for making personalized QoS prediction of cloud services. The framework provides an efficient global optimization scheme, thus offers robust and accurate prediction results. Also, it preserves explainability for QoS-prediction tasks which would be helpful for users make more definite selections.  The extensive experimental analysis indicates the effectiveness of our approach. \par 
	The main contribution of this paper lies in two-folds. First, we suggest learning neighborhood-based formal prediction model for personalized QoS evaluation of cloud services, and provide three distinctive variants of the model. Second, we prove that learning for neighborhood-based QoS prediction provides more robust and scalable method compared with traditional memory-based CF and matrix factorization techniques. \par
	Since neighborhood-based models are distinctly of matrix-factorization, we would like to integrate them together to have both of worlds in the future. 
	Resulting from the lack of real-world datasets for conducting experiments, we only conduct experimental studies on response-time, thus expect to adapt the proposed methods to the prediction tasks of other cloud service QoS properties, such as reliability, throughput and scalability. In addition, since we set out a general framework of the neighborhood model, extra information, such geo-information of users and items, temporal use of service invocation, can enter into this framework to offer more accurate prediction results.
	
	\begin{acknowledgements}
		The authors appreciate the reviewers for their informative comments. This work is supported by the Special Funds for Middle-aged and Young Core Instructor Training Programof Yunnan University,  the Applied Basic Research Project of Yunnan Province (2013FB009),  the Program for Innovative Research Team in Yunnan University (XT412011),  and the National Natural Science Foundation of China (61562090,61562092).
	\end{acknowledgements}
	
	\bibliographystyle{spmpsci}      
	\bibliography{mybib}   
\end{document}